\documentclass[utf8]{frontiersFPHY} 

\setcitestyle{square} 
\usepackage{url,hyperref,lineno,microtype,subcaption}
\usepackage[onehalfspacing]{setspace}

\def\keyFont{\fontsize{8}{11}\helveticabold }
\def\firstAuthorLast{A.S. Barabash} 
\def\Authors{A.S. Barabash\,$^{1,*}$}
\def\Address{$^{1}$ National Research Center "Kurchatov Institute", Institute of Theoretical and Experimental Physics, B. Cheremushkinskaya 25, 117218 Moscow, Russia }
\def\corrAuthor{A.S. Barabash}

\def\corrEmail{barabash@itep.ru}

\begin{document}
\onecolumn
\firstpage{1}

\title[Possibilities of future]{Possibilities of future double beta decay experiments to investigate inverted and normal ordering region of neutrino mass} 

\author[\firstAuthorLast ]{\Authors} 
\address{} 
\correspondance{} 

\extraAuth{}

\maketitle

\begin{abstract}
An overview of modern experiments on the search for neutrinoless double decay is presented. The obtained limits on the effective mass of the Majorana neutrino $\langle m_{\nu} \rangle$ are discussed taking into account the uncertainties in the value of the nuclear matrix elements (NMEs) and the value of the axial-vector constant $g_A$. Predictions for the values of $\langle m_{\nu} \rangle$ from the results of oscillation experiments and modern cosmological data are presented. The possibilities of the next generation experiments with sensitivity to $\langle m_{\nu} \rangle$ at the level of $\sim$ 10-50 meV (studying mainly the inverted ordering (IO) region) are discussed. 
The prospects for studying the normal ordering (NO)  region are discussed too. It is shown that the possibilities of studying the NO depend on the mass of the lightest neutrino m$_0$. In the limiting case of small mass (m$_0$ $\le$ 0.1 meV), the values of $\langle m_{\nu} \rangle$ $\approx$ 1-4 meV are predicted, which makes the study of this region inaccessible by the next generation experiments. But there is an allowed region of m$_0$ (7-30 meV) in the framework of NO, where the predicted values for $\langle m_{\nu} \rangle$ could be $\sim$ 10-30 meV and that is quite achievable for the next generation experiments. The possibility to rich in the future sensitivity to $\langle m_{\nu} \rangle$ at the level of $\sim$ 1-10 meV is also discussed. 

\section{}

\tiny
 \keyFont{ \section{Keywords:} neutrino mass, double beta decay, neutrino mass ordering, low background experiments} 
\end{abstract}

\section{Introduction}

The interest in neutrinoless double decay increased significantly after the discovery of neutrino oscillations in experiments with atmospheric, solar, reactor and accelerator neutrinos (see, for example, discussions in \cite{MOH06,BIL15,SAL18}). This is due to the fact that the very existence of neutrino oscillations indicates that the neutrino has a nonzero mass. However, oscillation experiments are not sensitive to the nature of the neutrino mass (Dirac or Majorana) and do not provide information on the absolute scale of neutrino masses. Registration of neutrinoless double beta decay will clarify many fundamental aspects of neutrino physics (see, for example, discussions in \cite{PAS06,BIL15a,VER16}):

 (i) lepton number non-conservation; 
 
 (ii) neutrino nature:  whether the neutrino is a Dirac or a Majorana particle;

 (iii) absolute neutrino mass scale; 
 
 (iv) the type of neutrino mass ordering (normal or inverted); 

(v) CP violation in the lepton sector (measurement of the Majorana CP-violating phases).

This process assumes a simple form, namely

\begin{equation}
(A,Z) \rightarrow (A,Z+2) + 2e^{-}.\label{eq:01}
\end{equation}

The discovery of this process is of fundamental interest, since it is practically the only way to establish the Majorana nature of neutrino. The Majorana nature of the neutrino would have interesting implications in many extensions of the Standard Model. For example the seesaw mechanism requires the existence of a Majorana neutrino to  explain the lightness of neutrino masses \cite{MIN77,YAN80,GEL79,MOH80}. A Majorana neutrino would also provide a natural explanation for the lepton number violation, and for the leptogenesis process which may explain the observed matter-antimatter asymmetry of the Universe \cite{FUK86}. 

The standard underlying 
mechanism behind neutrinoless double-beta decay is the exchange of a light Majorana neutrino. In this case, the half-life time of the decay can be presented as

\begin{equation}
   [T_{1/2}(0\nu)]^{-1} = G_{0\nu}g_A^4
   \mid{M_{0\nu}}\mid^2\Bigl|{\frac{\langle m_{\nu}\rangle}{m_e}}\Bigr|^2,\label{eq:02}
\end{equation}
where $G_{0\nu}$ is the phase space factor, which contains 
the kinematic information about the final state particles, 
and is exactly calculable to the precision of the input 
parameters \cite{KOT12,MIR15}, $g_A$ is the axial-vector coupling constant\footnote{Usually the value $g_A$ = 1.27 is used (the free neutron decay value). In nuclear matter, however, the value
of $g_A$ could be quenched. In $2\nu$ decay case effect of quenching could be quite strong \cite{VER16,BARE15,PIR15,KOS17,SUH17}. In case of $0\nu$ decay it can be a factor of $\sim$ 1.2-1.5 (see discussion in \cite{SUH17}). This question is still under discussion and there is no final answer up to now.},
$\mid{M_{0\nu}}\mid$ is the nuclear 
matrix element, $m_e$ is the mass of the electron, and 
$\langle m_{\nu}\rangle$ is the effective Majorana mass 
of the electron neutrino, which is defined as 
$\langle m_{\nu}\rangle$ = $\mid\sum_iU^2_{ei}m_i\mid$ where $m_i$ 
are the neutrino mass eigenstates and $U_{ei}$ 
are the elements of the neutrino mixing 
Pontecorvo-Maki-Nakagawa-Sakata (PMNS) matrix.

In contrast to two-neutrino decay (this decay has been detected - see review \cite{BAR15}, for example), neutrinoless double beta decay
has not yet been observed. The best limits on 
$\langle m_{\nu}\rangle$ are obtained for $^{136}$Xe, $^{76}$Ge, $^{130}$Te, $^{100}$Mo and $^{82}$Se (see Section 3). 
The assemblage of sensitive experiments
for different nuclei permits one to increase the reliability of the limit 
on $\langle m_{\nu}\rangle$. Present conservative limit can be set as 0.23 eV at 90\% C.L. 
(using conservative value from the KamLAND-Zen experiment). But one has to take into account that, in fact, this value could be in $\sim$ 1.5-2 times greater because of the possible quenching of $g_A$ (see recent discussions in \cite{SUH17}).

The main goal of next generation experiments is to investigate the IO region of neutrino mass ($\langle m_{\nu} \rangle$ $\approx$ (14-50) meV). If one will not see the decay in this region then it will be necessary to investigate region with $\langle m_{\nu} \rangle$  $<$ 14 meV.

\section{Predictions on $\langle m_{\nu} \rangle$ from neutrino oscillation and cosmological data}

Using the data of oscillatory experiments, one can obtain predictions for possible values of $\langle m_{\nu} \rangle$. Usually a so-called "lobster" ("crab") plot is constructed, which shows the possible values of $\langle m_{\nu} \rangle$, depending on the type of ordering and the mass of the lightest neutrino m$_0$, which is unknown (see, for example, recent papers \cite{VER16,PEN18,SAL18a}). The cosmological constraints on  $\Sigma m_{\nu}$ are used to limit the possible values of m$_0$. In Fig. 1, predictions on the effective Majorana neutrino mass are plotted as function of the lightest neutrino mass m$_0$. The 2$\sigma$ and 3$\sigma$ values of neutrino oscillation parameters are taken into account \cite{CAP17}. 
The PLANCK collaboration in a recent publication gives a limit of $\Sigma m_{\nu}$ $<$ 0.12 eV \cite{PLANCK}, using the new CMB data with different large scale structure observations. This leads to a limitation on m$_0$ $<$ 30 and $<$ 16 meV for normal and inverted ordering, respectively. Taking into account PLANCK's limit, different regions of possible values of $\langle m_{\nu} \rangle$ are obtained depending on the type of ordering:

1) $\langle m_{\nu} \rangle$ $\approx$ 14-50 meV for all values of  m$_0$ in the IO case.

2) In the NO case the situation is more complicated. The $\langle m_{\nu} \rangle$ can take values from practically 0 to 30 meV. And it has to be stressed that there is an allowed region of m$_0$ = 7-30 meV, where the $\langle m_{\nu} \rangle$ could be $\sim$ 10-30 meV and that is quite achievable for the next generation experiments. At m$_0$ = 10-30 meV, the NO and IO regions partially overlap and it will be difficult to uniquely determine the type of ordering. And only at m$_0$ $<$ 10 meV it will be possible to reliably distinguish between the NO and IO. At m$_0$ = 1-10 meV, a strong decrease in the values of $\langle m_{\nu} \rangle$ is possible for certain values of the Majorana phases (nevertheless, the probability of almost total nullification $\langle m_{\nu} \rangle$ is sufficiently small \cite{SAL18a}). At values of m$_0$ $\le$ 0.1 meV the $\langle m_{\nu} \rangle$ $\approx$ 1-4 meV (the so-called "limiting" case).

A global analysis of all available data was carried out in \cite{SAL18a} and it was shown that the NO is more preferable (at 3.5$\sigma$ level). It was also demonstrated that $\Sigma m_{\nu}$ $\ge$ 0.06 eV for the NO case, and $\Sigma m_{\nu}$ $\ge$ 0.1 eV for the IO. Nevertheless, the question of the order of the neutrino masses is not yet fully clarified and experiments on a double beta decay can contribute to its solution.  
A limit on $\langle m_{\nu} \rangle$ below 14 meV could be used to rule out the IO scheme, assuming that neutrinos are Majorana fermion. On the other hand a positive detection of $0\nu\beta\beta$ decay in the range that corresponds to $\langle m_{\nu} \rangle$ $>$ 14 meV would not give sufficient information to determine the mass ordering without an independent determination of m$_0$. Finally, in the context of three neutrino mixing, neutrinoless double beta decay experiments alone will be able to determine the neutrino mass ordering only ruling out the inverted scheme, that is to say if the ordering is normal and m$_0$ $\le$ 10 meV.

It is hoped that in a few years the value of $\Sigma m_{\nu}$  could be determined from cosmology (see, for example, discussions in \cite{SAL18a,BRI18}). This will help make a reliable conclusion about the type of ordering (for example, if the measured value will be less than 0.1 eV, it will mean that the NO is realized) and obtain information on the value of m$_0$. And this, in turn, will improve the predictions for a possible range of $\langle m_{\nu} \rangle$. For examlpe, in Ref. \cite{PEN18} it was demonstrated  that if the sum of neutrino masses is found to satisfy $\Sigma m_{\nu}$  $>$ 0.10 eV, then for NO case $\langle m_{\nu} \rangle$ $>$ 5 meV for
any values of the Majorana phases.

\section{Present status and current experiments}

Table 1 shows the best results for today on search for $0\nu\beta\beta$ decay for the most interesting nucleus-candidates for this process. Limits on the values of $T_{1/2}$ and $\langle m_{\nu} \rangle$ are given. To calculate $\langle m_{\nu}\rangle$ the NMEs from recent works \cite{BARE15,HYV15,SIM13,RAT13,ROD10,MEN09,NEA15,MUS13,SON17,TER15,FAN15,IWA16}
and the value $g_A$ = 1.27 have been used. One can see that the best modern experiments have reached a sensitivity of $\sim$ 10$^{25}$-10$^{26}$ years for the half-life and $\sim$ 0.1-0.3 eV for the $\langle m_{\nu} \rangle$. The spread in the values of the neutrino mass in each case is related to the currently existing uncertainties in the calculations of NMEs. Uncertainty in the values of NMEs is a factor of $\sim$ 2-3. As already noted, quenching of $g_A$ in the nucleus is possible and, as a result, the limits on the neutrino mass could be $\sim$ 1.5-2 times weaker. Table 1 shows that the most stringent limits on the effective mass of Majorana neutrino are obtained in experiments with $^{136}$Xe, $^{76}$Ge, $^{130}$Te, $^{100}$Mo and $^{82}$Se. For some nuclei, Table 1 lists two limit values for $T_{1/2}$ and $\langle m_{\nu} \rangle$. This is due to the fact that in some cases ($^{136}$Xe, $^{130}$Te and $^{76}$Ge) a large background fluctuation leads to too "optimistic" limits, substantially exceeding the "sensitivity" of the experiments. Therefore the values of the "sensitivity" of the experiments are also given in the Table 1. I believe that these values are although more conservative, but the most reliable. With this in mind, the conservative limit on $\langle m_{\nu} \rangle$ from modern double beta decay experiments is 0.23 eV (90\% C.L.).

Table 2 shows the best current and planned to start in 2018-2019 modern experiments that will determine the situation in the neutrinoless double beta decay in the coming years. 
It is seen that in the best of these experiments sensitivity to the $\langle m_{\nu} \rangle$ $\sim$ 0.04-0.2 eV will be achieved, which, apparently, will not be enough for verification  of the IO region (because to observe the effect, it is necessary to see the signal at least at 3$\sigma$ level; therefore, even the most sensitive experiments with the most favorable values of NMEs will not be able to register the decay).

\section{Possibilities of future double beta decay experiments to investigate IO region of neutrino mass}

Table 3 shows the most promising planned experiments, which will be realized in $\sim$ 5-15 years. To test the IO region of neutrino masses, it is necessary to achieve sensitivity to $\langle m_{\nu} \rangle$ at the level of $\sim$ 14-50 meV. 
Practically all experiments listed in Table 3 have a chance to register a $0\nu\beta\beta$  decay, but only  CUPID, nEXO and LEGEND-1000 overlap quite well the range of $\langle m_{\nu} \rangle$ associated with the IO. 
Thus, if the IO is actually realized in nature and the neutrino is Majorana particle, then it is likely that the neutrinoless double beta decay will be registered in the experiment in $\sim$ 5-15 years. And the CUPID, nEXO and LEGEND-1000 experiments have  the greatest chances to see the effect. But even these, the most sensitive experiments, do not guarantee the observation of the effect. At unfavorable values of NMEs and  $g_A$ , the sensitivity of these experiments will be insufficient to completely cover the entire range of possible values of  $\langle m_{\nu} \rangle$ for the IO. And one has to remember that in order to observe the effect it is necessary to have at least 3$\sigma$ confidence level (in Table 3, the sensitivity is indicated at 90\% C.L. (1.6$\sigma$)).

\section{Possibilities of future double beta decay experiments to investigate NO region of neutrino mass}

In the NO case, the following possible ranges of $\langle m_{\nu} \rangle$ can be distinguished:

1) 10-30 meV. In this case, $0\nu\beta\beta$ decay could be detected in the next generation experiments (see Table 3). But, for this area of mass, it will be difficult to distinguish the NO from IO. In this case additional information about m$_0$ is required.

2) 3-10 meV. In this case, detectors containing $\sim$ 1-10 tons of $\beta\beta$ isotope are  required.  And it is possible (in principle) to investigate this region of $\langle m_{\nu} \rangle$ in the future (sensitivity to $T_{1/2}$ on the level of $\sim$ $10^{28}-10^{29}$ yr will be needed).

3) 1-3 meV. In this case, detectors containing $\sim$ 10-100 tons of $\beta\beta$ isotope are  required. It will be very difficult (if possible) to investigate this region of $\langle m_{\nu} \rangle$ in the future (sensitivity to $T_{1/2}$ on the level of $\sim$ $10^{29}-10^{30}$ yr will be needed).

4) $<$ 1 meV. This area is not available for observation in foreseeable future.

The possibility of studying $0\nu\beta\beta$ decay with sensitivity to neutrino mass on the level of $\sim$ 1-5 meV has been analysed in \cite{BAR18}. It was shown that the 3-5 meV region can be studied by detectors containing $\sim$ 10 tons of $\beta\beta$ isotope. Moreover, the detectors should have a sufficiently high efficiency ($\sim$ 100\%), good energy resolution (FWHM $<$1-2\%), and low level of background in the investigated region ($\sim$ $10^{-6}-10^{-7}$ c/kev$\times$kg$\times$yr). In addition, the cost of an isotope becomes important and can seriously limit the feasibility of such experiments \cite{BAR18}. It was noted in \cite{BAR18} that $^{136}$Xe, $^{130}$Te,  $^{82}$Se, $^{100}$Mo and $^{76}$Ge are most promising isotopes, and the most suitable experimental techniques are low-temperature scintillation bolometers, gas Xe TPC and HPGe semiconductor detectors.

Summarizing all of the above, one can conclude that if we are dealing with the NO and $\langle m_{\nu} \rangle$ = 10-30 meV, then $0\nu\beta\beta$ decay could be registered in next-generation experiments ($\sim$ 5-15 years from now). To study the range of $\langle m_{\nu} \rangle$ $<$10 meV, new, more sensitive experiments with the mass of the investigated isotope $\sim$ 1-10 tons ($\langle m_{\nu} \rangle$ = 3-10 meV) or $\sim$ 10-100 tons ($\langle m_{\nu} \rangle$ = 1-3 meV) are required. In more detail, such possible experiments are discussed in \cite{BAR18}.

\section{Conclusion}

Thus, we can conclude that the present conservative limit on $\langle m_{\nu} \rangle$ from  double beta decay experiments is 0.23 eV (90\% C.L.). Within the next 3-5 years, the sensitivity of modern experiments will be brought to $\sim$ 0.04-0.2 eV. To study the IO region  (0.014-0.05 eV), new generation experiments will be realised, which will achieve the required sensitivity in $\sim$ 5-15 years. If we are dealing with NO, then everything depends on the value of  $\langle m_{\nu} \rangle$ that is realized in nature. If $\langle m_{\nu} \rangle$ = 10-30 meV, then this lies in the sensitivity region of the next generation experiments and $0\nu\beta\beta$ decay could be registered. If $\langle m_{\nu} \rangle$ = 3-10 meV, new, more sensitive experiments with $\sim$ 1-10 tons of $\beta\beta$ isotope are  required (and it seems possible). For $\langle m_{\nu} \rangle$ = 1-3 meV experiments with of $\sim$ 10-100 tons of the isotope are required and it will be very difficult (if possible) to reach needed sensitivity in this case. If, however, $\langle m_{\nu} \rangle$ $\le$1 meV, then apparently $0\nu\beta\beta$ decay will not be registered in the foreseeable future\footnote{In this article, we proceed from the assumption that we are dealing with an  $0\nu\beta\beta$ decay, going through the exchange of light Majorana neutrinos. This mechanism is the most popular at the moment. Nevertheless, it should be emphasized that, in principle, other mechanisms are possible (right-handed currents, supersymmetry, heavy neutrinos, doubly charged Higgs bosons, etc.) - see, for example, discussions in \cite{VER16,ENG17,AVI08}. Therefore, if the $0\nu\beta\beta$ decay will be detected, then, first of all, it will be necessary to verify that we are dealing with a mechanism associated with a light neutrino. And only after that it will be possible to make a reliable conclusion about the value of $\langle m_{\nu} \rangle$.
It is not excluded that several mechanisms will contribute to the $0\nu\beta\beta$ transition at the same time. In this case, it will be difficult to determine the true value of $\langle m_{\nu} \rangle$. On the other hand, the presence of other decay mechanisms allows us to hope for the registration of $0\nu\beta\beta$ decay even at very low value of $\langle m_{\nu} \rangle$.}.



\section*{Author Contributions}

The author confirms being the sole contributor of this work and
approved it for publication.


\section*{Acknowledgments}
This work was supported by Russian Science Foundation (grant No. 18-12-00003).
And I want to thank Prof. F. Simkovic and V. Umatov for help in preparing the figure.

\bibliographystyle{frontiersinHLTH&FPHY} 
\bibliography{test}

\begin{thebibliography}{0} 
\bibitem{MOH06}
 Mohapatra RN, Smirnov AY. 
 Neutrino mass and new physics. {\it Annu. Rev. Nucl. Part. Sci.} (2006) 56:569-628. DOI: 10.1146/annurev.nucl.56.080805.140534
\bibitem{BIL15}
Bilenky SM.  
Neutrino in Standard Model and beyond. {\it Phys. Part. Nucl.} (2015) 46:475-496. DOI: 10.1134/S1063779615040024
\bibitem{SAL18}
de Salas PF et al.  
Status of neutrino oscillations 2018: first hint for normal mass ordering and improved CP sensitivity. {\it Phys. Let. B} (2018) 782:633-640. DOI: 10.1016/j.physletb.2018.06.019 
\bibitem{PAS06}
Pascoli S, Petcov ST, Schwetz T.  
The absolute neutrino mass scale, neutrino mass
spectrum, Majorana CP-violation and neutrinoless double-beta decay. {\it Nucl. Phys. B} (2006) 734: 24-49. DOI: 10.1016/j.nuclphysb.2005.11.003
\bibitem{BIL15a}
Bilenky SM, Giunti C. 
Neutrinoless double-beta decay: a probe of physics beyond the Standard Mode. {\it Int. J. Mod. Phys. A} (2015) 30:1530001. DOI: 10.1142/S0217751X1530001X
\bibitem{VER16}
Vergados JD, Ejiri H, Simkovic F.  
Neutrinoless double beta decay and neutrino mass. {\it Int. J. Mod. Phys. E} (2016) 25:163007. DOI: 10.1142/S0218301316300071
\bibitem{MIN77}
Minkowski P.
$\mu\to$e$\gamma$ at a rate of one out of 10$^9$ muon decays? {\it Phys. Lett. B} (1977) 67:421-428. DOI: 10.1016/0370-2693(77)90435-X
\bibitem{YAN80}
Yanagida T. 
Horizontal symmetry and masses of neutrinos. {\it Prog. Theor. Phys.} (1980) 64:1103-1105. DOI: 10.1143/PTP.64.1103
\bibitem{GEL79}
Gell-Mann PRM, Ramond P, Slansky R. 
In Proceedings of the
{\it Supergravity Stony Brook Workshop}, p.315, New York (North Holland Publishing Co.) (1979).
\bibitem{MOH80}
Mohapatra RN, Senjanovic G.   
Neutrino mass and spontaneous parity nonconservation. {\it Phys. Rev. Lett.} (1980) 44:912-915. DOI: 10.1103/PhysRevLett.44.912
\bibitem{FUK86}
Fukugita M, Yanagida T.  
Barygenesis without grand unification. {\it Phys. Lett. B} (1986) 174:45-47. DOI: 10.1016/0370-2693(86)91126-3
\bibitem{KOT12}
Kotila J, Iachello F.
Phase space factors for double-$\beta$ decay. {\it Phys. Rev. C} (2012) 85:034316. DOI: 10.1103/PhysRevC.85.034316
\bibitem{MIR15}
Mirea M, Pahomi T, Stoica S. 
Values of the phase space factors involved in double beta decay.  {\it Rom. Rep. Phys.} (2015) 67: 872-889.
\bibitem{BARE15}
Barea J, Kotila J, Iachello F.  
$0\nu\beta\beta$ and $2\nu\beta\beta$ nuclear matrix elements in the interacting boson model with isospin restoration. {\it Phys. Rev. C} (2015) 91:034304. DOI: 10.1103/PhysRevC.91.034304
\bibitem{PIR15}
Pirinen P, Suhonen J.  
Systematic approach to $\beta$ and $2\nu\beta\beta$  decays of mass A=100–136 nuclei. {\it Phys. Rev. C} (2015) 91:054309. DOI: 10.1103/PhysRevC.91.054309
\bibitem{KOS17}
Kostensalo J, Haaranen M, Suhonen J.  
Electron spectra in forbidden $\beta$ decays and the quenching of the weak axial-vector coupling constant g$_A$. {\it Phys. Rev. C} (2017) 95:044313. DOI: 10.1103/PhysRevC.95.044313
\bibitem{SUH17}
Suhonen J.  
Impact of the quenching of g$_A$ on the sensitivity of $0\nu\beta\beta$ experiments. {\it Phys. Rev. C} (2017) 96:055501. DOI: 10.1103/PhysRevC.96.055501
\bibitem{BAR15}
Barabash AS.  
Average and recommended half-life values for two-neutrino double beta decay. {\it Nucl. Phys. A} (2015) 935:52-64. DOI: 10.1016/j.nuclphysa.2015.01.001
\bibitem{PEN18}
Penedo JT, Petcov ST.  
The 10$^{-3}$ eV frontier in neutrinoless double beta decay. {\it Phys. Lett. B} (2018) 786:410-417. DOI: 10.1016/j.physletb.2018.09.059 
\bibitem{SAL18a}
de Salas PF et al.  
Neutrino mass ordering from oscillations and beyond: 2018 status and future prospects.
{\it Front. Astron. Space. Sci.} (2018) 5:00036. DOI: 10.3389/fspas.2018.00036
\bibitem{CAP17}
Capozzi F et al.
Global constraints on absolute neutrino masses and their ordering. {\it Phys. Rev. D} (2017) 95:096014. DOI: 10.1103/PhysRevD.95.096014 
\bibitem{BRI18}
Brinckmann T et al. The promising future of a robust cosmological neutrino mass measurement.
arXiv:1808.05955 [astro-ph.CO]. 

\bibitem{PLANCK}
 Planck Collaboration: Aghanim N et al.
 Planck 2018 results. VI. Cosmological parameters. arXiv:1807.06209 [astro-ph.CO].
\bibitem{HYV15}
Hyvarinen J, Suhonen J. 
Nuclear matrix elements for $0\nu\beta\beta$ decays with light or heavy Majorana-neutrino exchange. {\it Phys. Rev. C} (2015) 91:024613. DOI: 10.1103/PhysRevC.91.024613
\bibitem{SIM13}
Simkovic F, Rodin V,  Faessler A, Vogel P.  
$0\nu\beta\beta$ and $2\nu\beta\beta$ nuclear matrix elements, QRPA, and isospin symmetry restoration. {\it Phys. Rev. C} (2013) 87:045501. DOI: 10.1103/PhysRevC.87.045501
\bibitem{RAT13}
Rath PK et al. 
Neutrinoless $\beta\beta$ decay transition matrix elements within mechanisms involving light Majorana neutrinos, classical Majorons and sterile neutrinos. {\it Phys. Rev. C} (2013) 88:064322. DOI: 10.1103/PhysRevC.88.064322
\bibitem{ROD10}
Rodriguez TR, Martinez-Pinedo G.  
Energy density functional study of nuclear matrix elements for neutrinoless $\beta\beta$ decay. {\it  Phys. Rev. Lett.} (2010) 105:252503. DOI: 10.1103/PhysRevLett.105.252503
\bibitem{MEN09}
Menendez J, Poves A, Caurier E, Nowacki F. 
Disassembling the nuclear matrix elements of the neutrinoless $\beta\beta$ decay. {\it Nucl. Phys. A} (2009) 818:139-151. DOI: 10.1016/j.nuclphysa.2008.12.005
\bibitem{NEA15}
Neacsu A, Horoi M. 
Shell model studies of the $^{130}$Te neutrinoless double-$\beta$ decay
 {\it Phys. Rev. C} (2015) 91:024309. DOI: 10.1103/PhysRevC.91.024309
\bibitem{MUS13}
Mustonen M, and Engel J.  
Large-scale calculations of the double-$\beta$ decay of $^{76}$Ge, $^{130}$Te, $^{136}$Xe, and $^{150}$Nd in the deformed self-consistent Skyrme quasiparticle random-phase approximation. {\it Phys. Rev. C} (2013) 87:064302. DOI: 10.1103/PhysRevC.87.064302
\bibitem{SON17}
Song LS et al.  
Nuclear matrix element of neutrinoless double-$\beta$ decay: Relativity and short-range correlations. {\it Phys. Rev. C} (2017) 95:024305. DOI: 10.1103/PhysRevC.95.024305
\bibitem{TER15}
Terasaki J.  
Many-body correlations of quasiparticle random-phase approximation in nuclear matrix elements of neutrinoless double-$\beta$ decay. {\it Phys. Rev. C} (2015) 91:034318. DOI: 10.1103/PhysRevC.91.034318
\bibitem{FAN15}
 Fang D-L, Faessler A, Simkovic F.  
 Partial restoration of isospin symmetry for neutrinoless double $\beta$ decay in the deformed nuclear system of $^{150}$Nd. {\it Phys. Rev. C} (2015) 92:044301. DOI: 10.1103/PhysRevC.92.044301
\bibitem{IWA16}
Iwata Y et al.  
Large-scale shell-model analysis of the neutrinoless $\beta\beta$ decay of $^{48}$Ca. {\it Phys. Rev. Lett.} (2016) 116:112502. DOI: 10.1103/PhysRevLett.116.112502
\bibitem{UME08}
Umehara S.  
Neutrino-less double beta decay of $^{48}$Ca studied by CaF$_2$ (Eu) scintillators. {\it Phys. Rev. C} (2008) 78:058501. DOI: 10.1103/PhysRevC.78.058501
\bibitem{AGO18}
Agostini M et al.  
Improved limit on neutrinoless double-$\beta$ decay of $^{76}$Ge from GERDA Phase II. {\it Phys. Rev. Lett.} (2018) 120:132503. DOI: 10.1103/PhysRevLett.120.132503
\bibitem{AZZ18}
Azzolini O et al. 
First result on the neutrinoless double beta decay of $^{82}$Se with CUPID-0. {\it Phys. Rev. Lett.} (2018) 120:232502. DOI: 10.1103/PhysRevLett.120.232502
\bibitem{ARN10}
Arnold R et al. 
Measurement of the two neutrino double beta decay half-life of Zr-96 with the NEMO-3 detector. {\it Nucl. Phys. A} (2010) 847:168-179. DOI: 10.1016/j.nuclphysa.2010.07.009
\bibitem{ARN15}
Arnold R et al.  
Results of the search for neutrinoless double-$\beta$ decay in $^{100}$Mo
with the NEMO-3 experiment. {\it Phys. Rev. D} (2015) 92:072011. DOI: 10.1103/PhysRevD.92.072011
\bibitem{BAR18a}
Barabash AS et al.  
Final results of the Aurora experiment to study 2$\beta$ decay
of $^{116}$Cd with enriched $^{116}$CdWO$_4$ crystal scintillators. {\it Phys. Rev. D} (2018) 98:092007. DOI: 10.1103/PhysRevD.98.092007 
\bibitem{ALD18}
Alduino C et al.  
First results from CUORE: a search for lepton number violation
via $0\nu\beta\beta$ decay of $^{130}$Te. {\it Phys. Rev. Lett.} (2018) 120:132501. DOI: 10.1103/PhysRevLett.120.132501
\bibitem{GAN16}
 Gando A et al.  
 Search for Majorana neutrinos near the inverted mass hierarchy
region with KamLAND-Zen. {\it Phys. Rev. Lett.} (2016) 117:082503.  DOI:  10.1103/PhysRevLett.117.082503
\bibitem{ARN16}
Arnold R et al.  
Measurement of the $2\nu\beta\beta$ decay half-life of $^{150}$Nd and a search for $0\nu\beta\beta$ decay processes with the full exposure from the NEMO-3 detector.  {\it Phys. Rev. D} (2016) 94: 072003. DOI: 10.1103/PhysRevD.94.072003
\bibitem{AAL18}
Aalseth CE et al. 
Search for neutrinoless double-$\beta$ decay in $^{76}$Ge with the MAJORANA DEMONSTRATOR. {\it Phys. Rev. Lett.} (2018) 120:132502.  DOI: 10.1016/0370-2693(77)90435-X
\bibitem{ALB18}
Albert JB et al.  
Search for neutrinoless double-beta decay with the upgraded EXO-200 detector. {\it Phys. Rev. Lett.} (2018) 120:072701. DOI: 10.1103/PhysRevD.97.072007
\bibitem{SHI18}
Shirai J.  
KamLAND-Zen. In Proceedings of XVII International Workshop on Neutrino Telescopes, 13-17 March 2017, Venezia, Italy (Pros. Science (Neutel2017) (2018) p. 027). DOI: 10.22323/1.307.0027
\bibitem{AND16}
Andringa S et al.  
Current status and future prospects of the SNO+ experiment. {\it Adv. High Energy Phys.} (2016) 2016: 6194250. DOI: 10.1155/2016/6194250
\bibitem{FIS18}
 Fischer V.  
 Search for neutrinoless double-beta decay with SNO+. arXiv:1809.05986 [physics.ins-det].
\bibitem{MAR16}
Martin-Albo J et al.   
Sensitivity of NEXT-100 to neutrinoless double
beta decay. JHEP (2016) 5:159. DOI: 10.1007/JHEP05(2016)159
\bibitem{POD17}
Poda DV. 
$^{100}$Mo-enriched Li$_2$MoO$_4$ scintillating bolometers for $0\nu\beta\beta$ decay search: from LUMINEU to CUPID-0/Mo projects. {\it AIP Conf. Proc.} (2017) 1894:020017. DOI: 10.1063/1.5007642
\bibitem{YOO18}
Yoon YS.  
Status of the AMoRE experiment to search for neutrinoless double beta decay of Mo-100. PoS (ICRC2017) (2018) 1056. DOI: 10.22323/1.301.1056
\bibitem{KIM18}
Kim SC.  
The Status of the AMoRE experiment. {\it In: Proc. Neutrino 2018 – XXVIII Int. Conf. Physics and Astrophysics.} (2018). Zenodo. http://doi.org/10.5281/zenodo.1300715


\bibitem{BAR17}
Barabash AS et al.  
Calorimeter development for the SuperNEMO double beta decay
experiment. {\it Nucl. Instrum. Methods A} (2017) 868:98-108. DOI: 10.1016/j.nima.2017.06.044
\bibitem{CAS16}
Cascella M et al.  
Construction and commissioning of the SuperNEMO detector tracker. {\it Nucl. Instrum. Methods A} (2016) 824:507-509. DOI: 10.1016/j.nima.2015.11.083 
\bibitem{ABG17}
Abgrall N et al.  
The large enriched germanium experiment for neutrinoless double beta decay (LEGEND). {\it AIP Conf. Proc.} (2017) 1894:02027. DOI: 10.1063/1.5007652
\bibitem{ALB18a}
Albert JB et al.  
Sensitivity and discovery potential of the proposed nEXO experiment
to neutrinoless double-$\beta$ decay. {\it Phys. Rev. C} (2018) 97:065503. DOI: 10.1088/1748-0221/13/01/P0
\bibitem{WAN15}
Wang G et al. 
CUPID: CUORE (Cryogenic Underground Observatory for Rare Events)
Upgrade with Particle IDentification. arXiv:1504.03599 [physics.ins-det].
\bibitem{WAN15a}
 Wang G et al.  
 R$\&$D towards CUPID ( CUORE Upgrade with Particle IDentification). arXiv:1504.03612 [physics.ins-det].
\bibitem{CHE17}
Chen X et al.
 PandaX-III: Searching for neutrinoless double beta decay
with high pressure $^{136}$Xe gas time projection chambers. {\it Sci. China Phys. Mech. Astron.} (2017) 60:061011. DOI: 10.1007/s11433-017-9028-0
\bibitem{BAR18}
Barabash AS.  
Main features of detectors and isotopes to investigate double beta decay with increased sensitivity. {\it Int. J. Mod. Phys. A} (2018) 33:1843001. DOI: 10.1142/S0217751X18430017
\bibitem{ENG17}
Engel J, Menendez J.
Status and future of nuclear matrix elements for neutrinoless double-beta decay: a review.
{\it Rep. Prog. Phys.} (2017) 80:046301. DOI: 10.1088/1361-6633/aa5bc5
\bibitem{AVI08}
Avignone FT, Elliott SR, Engel J.
Double beta decay, Majorana neutrinos, and neutrino mass.
{\it Rev Mod. Phys.} (2008) 80:481-516. DOI: 10.1103/RevModPhys.80.481  

\end{thebibliography}

\begin{table}[ph]
\caption{Best present limits on $0\nu\beta\beta$ decay (at
90\% C.L.). To calculate $\langle m_{\nu}\rangle$ 
the NME from \cite{BARE15,HYV15,SIM13,RAT13,ROD10,MEN09,NEA15,MUS13,SON17}, 
phase-space factors from \cite{KOT12,MIR15} and $g_A$ = 1.27 have been used.
In case of $^{150}$Nd NME from \cite{TER15,FAN15} 
and in case of $^{48}$Ca from \cite{IWA16} were used in addition. The bold type denotes the so-cold “sensitivity” values (see text).}
{\begin{tabular}{@{}cccccc@{}} \toprule
Isotope & Q$_{2\beta}$, keV & $T_{1/2}$, yr & $\langle m_{\nu} \rangle$, eV 
 & Experiment & References \\ 

\hline
$^{48}$Ca & 4267.98  & $>5.8\times10^{22}$ & $<3.1-15.4$ & CANDLES & \cite{UME08} \\
$^{76}$Ge & 2039.00 & ${\bf >5.8\times10^{25}}$ & ${\bf <0.14-0.37}$ & GERDA-I+GERDA-II & \cite{AGO18} \\
& & ($>8\times10^{25}$) & ($<0.12-0.31$) & & \\
$^{82}$Se &2997.9 & $>2.4\times10^{24}$ & $<0.4-0.9$ & CUPID-0/Se & \cite{AZZ18} \\
$^{96}$Zr & 3355.85 & $>9.2\times10^{21}$ & $<3.6-10.4$ & NEMO-3 & \cite{ARN10} \\
$^{100}$Mo & 3034.40 & $>1.1\times10^{24}$ & $<0.33-0.62$ & NEMO-
3 & \cite{ARN15} \\
$^{116}$Cd & 2813.50 & $>2.2\times10^{23}$ & $<1-1.7$ & AURORA & \cite{BAR18a} \\
$^{128}$Te & 866.6 & $>1.5\times10^{24}$ & $2.3-4.6$ & Geochem. exp. &(see \cite{BAR15})  \\
$^{130}$Te & 2527.52 & ${\bf >7\times10^{24}}$ & ${\bf <0.19-0.74}$ & CUORICINO + & \\
&  & ($>1.5\times10^{25}$) & ($<0.13-0.50$) & CUORE0 + CUORE & \cite{ALD18} \\
$^{136}$Xe & 2457.83 & ${\bf >5.6\times10^{25}}$ & ${\bf <0.08-0.23}$ & KamLAND-Zen
&  \cite{GAN16} \\
 & & ($>1.07\times10^{26}$) & ($<0.06-0.16$) & &\\
$^{150}$Nd & 3371.38 & $>2\times10^{22}$ & $<1.6-5.3$ & NEMO-3 & \cite{ARN16} \\ \botrule
\end{tabular} \label{ta1}}
\end{table}

\begin{table}[ph]
\caption{Best current and planned to start in 2018-2019 modern experiments. 
Sensitivity at 90\% C.L. for three (GERDA-II, Majorana Demonstrator, 
SuperNEMO Demonstrator and KamLAND-Zen) and
five (for other experiments)  
years of measurements is presented. M is mass of the isotope.}
{\begin{tabular}{@{}ccccccc@{}} \toprule
Experiment & Isotope & M, kg & Sensitivity & Sensitivity & Status & References \\
& &  & $T_{1/2}$, yr & $\langle m_{\nu} \rangle$, meV & &  \\ 
\hline
CUORE  & $^{130}$Te & 200 & $9.5\times10^{25}$ & 53--200 & current & \cite{ALD18} \\  
GERDA-II & $^{76}$Ge & 35 & $1.5\times10^{26}$ & 90--230 & current & \cite{AGO18} \\ 
Majorana-D  & $^{76}$Ge & 30 & $1.5\times10^{26}$ & 90--230 & current & \cite{AAL18} \\
EXO-200 & $^{136}$Xe & 200 & $5.7\times10^{25}$ & 85--225 & current & \cite{ALB18} \\
CUPID-0/Se & $^{82}$Se & 5 & 6$\times10^{24}$ & 250--590 & current & \cite{AZZ18} \\
KamLAND-Zen  & $^{136}$Xe & 750 & 2$\times10^{26}$ & 45--120 & start in 2018 & \cite{SHI18} \\
SNO+-I & $^{130}$Te & 1300 & $2\times10^{26}$ & 36--140 & start in 2019 & \cite{AND16,FIS18}  \\
NEXT & $^{136}$Xe & 100 & $6\times10^{25}$ & 83--220 & start in 2019 & \cite{MAR16} \\ 

CUPID-0/Mo & $^{100}$Mo & 4 & 1.5$\times10^{25}$ & 90--170 & start in 2019 & \cite{POD17} \\
AMoRE-I & $^{100}$Mo & 2.5 & $\sim 10^{25}$ & 110--210 & start in 2019 & \cite{YOO18,KIM18} \\
SuperNEMO-D & $^{82}$Se & 7 & 6.5$\times10^{24}$ & 240--560 & start in 2019 & \cite{BAR17,CAS16} \\
 \botrule
\end{tabular}\label{ta2}}
\end{table}

\begin{table}[ph]
\caption{Main most developed and promising projects for next generation experiments. 
Sensitivity at 90\% C.L. for five (KamLAND2-Zen, SNO+-II, AMoRE-II, SuperNEMO, PandaX-III, LEGEND-200) and ten (LEGEND-1000, nEXO and CUPID) years of measurements is presented. M is mass of the isotope.}
{\begin{tabular}{@{}ccccccc@{}} \toprule
Experiment & Isotope & M, kg & Sensitivity & Sensitivity & Status & References \\
& &  & $T_{1/2}$, yr & $\langle m_{\nu} \rangle$, meV & &  \\ 
\hline
LEGEND & $^{76}$Ge & 200 & $\sim 10^{27}$ & 34--90 & in progress & \cite{ABG17} \\
       &  & 1000 & $\sim 10^{28}$ & 11-28 & R\&D &  \\
nEXO & $^{136}$Xe & 5000 & $9\times 10^{27}$ & 8--22 & R\&D & \cite{ALB18a} \\ 
CUPID & $^{130}$Te, $^{100}$Mo & $\sim$ 200-500 & $(2-5)\times10^{27}$ & 6-17 & R\&D & \cite{WAN15,WAN15a} \\
& $^{82}$Se, $^{116}$Cd &  &  &  &  & \\
KamLAND2-Zen & $^{136}$Xe & 1000 &  $6\times10^{26}$ & 25-70 & R\&D & \cite{SHI18} \\
SNO+-II & $^{130}$Te & 8000 & $7\times 10^{26}$ & 20-70 &  R\&D & \cite{AND16} \\ 
AMoRE-II & $^{100}$Mo & 100 & $5\times10^{26}$ & 15-30 & R\&D & \cite{KIM18} \\
SuperNEMO & $^{82}$Se & 100--140 & $(1-1.5)\times10^{26}$ & 50--140 & R\&D & \cite{BAR17,CAS16} \\
PandaX-III & $^{136}$Xe & 200 & $\sim 10^{26}$ & 65--170 & R\&D & \cite{CHE17} \\ 
  &  & 1000 &  $\sim 10^{27}$ & 20-55 & R\&D & \\ 
 \botrule
\end{tabular}\label{ta2}}
\end{table}

\begin{figure}[h!]
\begin{center}
\includegraphics[width=10cm]{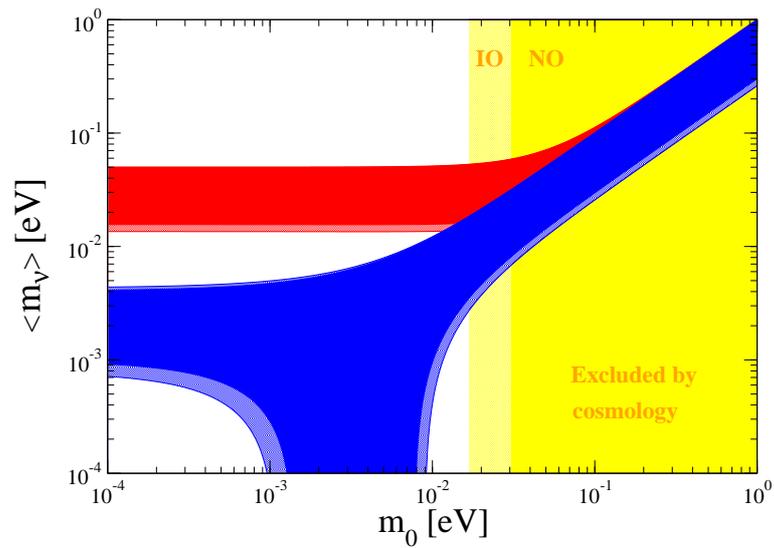}
\end{center}
\caption{Predictions on $\langle m_{\nu} \rangle$ from neutrino oscillations versus the lightest neutrino mass m$_0$ in the two cases of normal (the blue region) and inverted (the red region) spectra. The 2$\sigma$ and 3$\sigma$ values of neutrino oscillation parameters are considered \cite{CAP17}. The excluded region by cosmological data ($\Sigma m_{\nu}$ $<$ 0.12 eV) m$_0$ is  presented in yellow ($>$ 30 meV for the NO and $>$ 16 meV for the IO).}\label{fig:1}
\end{figure}





\end{document}